\documentclass[aps,preprint,showpacs,amsmath,amssymb,prd,nofootinbib]{revtex4}
\usepackage{amssymb,graphics,graphicx}
\newcommand{\be}{\begin{equation}} \newcommand{\ee}{\end{equation}}

\begin{document}
\title
{\Large \bf Quartification with $T^{'}$ Flavor}
\author{David A. Eby$^{1}$\footnote{daeby@physics.unc.edu}, Paul H. Frampton$^{1}$\footnote{paul.h.frampton@gmail.com}, Xiao-Gang He$^{2,3}$\footnote{hexg@phys.ntu.edu.tw}, and Thomas W. Kephart$^{4}$\footnote{tom.kephart@gmail.com}}

\affiliation{$^{1}$Department of Physics and Astronomy, University of North Carolina, 
Chapel Hill, NC 27599-3255, USA\\
$^{2}$Department of Physics and Center for Theoretical Sciences, National Taiwan University, Taipei, Taiwan\\
$^3$INPAC, Department of Physics, Shanghai Jiao Tong University, Shanghai, China\\
$^{4}$Department of Physics and Astronomy, Vanderbilt University, Nashville, Tennessee 37235, USA}

\date{\today}

\begin{abstract}
In the simplest (non-quiver) unified theories, fermion families are often
treated sequentially and a flavor symmetry may act similarly. As an
alternative with non-sequential flavor symmetry, we 
consider a model based on the group
$(T^{'} \times Z_2)_{global} \times [SU(3)^4]_{local}$
which combines the predictions of $T^{'}$ flavor
symmetry with the features of a unified quiver gauge theory.
The model accommodates the relationships between mixing angles
separately for neutrinos, and for quarks, which have been
previously predicted with $T^{'}$. This quiver unification theory 
makes predictions of
several additional gauge bosons 
and bifundamental fermions at the TeV scale.

\noindent

\end{abstract}

\maketitle

\section{Introduction}
In order to address the question of masses and mixing angles which
occur for quarks and leptons in the standard model, one promising
direction is to introduce a flavor symmetry that commutes
with the standard model gauge group. By judicious assignments of
the particles to specific representations of the flavor symmetry,
one can obtain relations between parameters in the model.
The flavor symmetry may treat the fermion families differently so that the 
simplest approaches to gauge unification are inapplicable. The present 
article will show how to combine the flavor group ($T^{'}$),
which has been studied previously
\cite{Frampton:1994rk,EFM1,Aranda:2000tm,Frampton:2007et,Feruglio:2007uu,Frampton:2008bz,Chen:2007t1,Chen:2009t2}, 
with a quiver unified quartification $SU(3)^4$ gauge group~\cite{Babu:2007fx}, while successfully keeping
results previously obtained without unification, such
as the Cabibbo angle~\cite{Frampton:2008bz}, as well as tribimaximal mixing for neutrinos 
\cite{Ma:2001dn,HPS,He:2003rm,Babu:2005se,
Lee:2006pr,Frampton:2008ci}. The quiver unification has the
advantage of implying further relationships
between the gauge couplings.

\bigskip

\section{The model}

We first consider a quartification $(SU(3)^4)$ model with bifundamental chiral fermions
in the usual arrangement of bifundamentals, 
but find we can not make the necessary charge assignments to recover 
the requisite $T^{'}$ family symmetry. This will lead us to add a sub-quiver
of fermions to accommodate $T^{'}$ quartification.

\bigskip

\noindent
Quartification, from its inception by Joshi and Volkas~\cite{Joshi:1992qu}, has 
historically been used for gauge-coupled unification without supersymmetry and for leptonic 
color models~\cite{Babu:2004co,Demaria:2005in,Chen:2004qr1,Demaria:2006qr2,Demaria:2007qr3,Babu:2007fx}.  
Many of these models have adapted the
same unification techniques as the first GUT theories~\cite{G:1974gut}. There have 
been several significant milestones in this approach (and several different preferred unification
scales) including partial unification~\cite{Joshi:1992qu}, 
complete unification~\cite{Babu:2004co}, and intermediate symmetry breaking~\cite{Demaria:2005in}.
We choose a different style of unification compared with prior work on quartification, one 
predicated upon the mechanism in Refs.~\cite{Frampton:2002un1,Frampton:2003un2}, 
that by embedding

\be
SU(3)_C \times SU(2)_L \times U(1)_Y \ ,
\ee

\noindent
in $SU(3)^N$ we naturally achieve unification in the TeV region. This is accomplished by replacing the logarithmic evolution of couplings, with the use of group theoretic factors.

\bigskip

\noindent
The quartification gauge group is the quasi-simple

\be
SU(3)_C \times SU(3)_L \times SU(3)_{\ell} \times SU(3)_R \ ,
\ee

\bigskip

\noindent
with couplings equal up to numerical group theory factors~\cite{Frampton:2002un1,Frampton:2003un2}.
Let the family symmetry be
\be
 T' \times Z_2\  ,
\ee
with the minimal anomaly-free bifundamental chiral fermions:
\be 
3[ (3,\bar{3},1,1) +   (\bar{3},1,1,3) 
 +(1,3,\bar{3},1) +(1,1,3,\bar{3})] \ .
\ee
We shall assign the leptons to
irreps as follows \cite{Frampton:2008bz}~:
\be
\begin{array}{ccc}
\left. \begin{array}{c}
(13\bar3 1)_3\supset\left( \begin{array}{c} \nu_{\tau} \\ \tau^- \end{array} \right)_{L} \\
(13\bar3 1)_2\supset\left( \begin{array}{c} \nu_{\mu} \\ \mu^- \end{array} \right)_{L} \\
(13\bar3 1)_1\supset\left( \begin{array}{c} \nu_e \\ e^- \end{array} \right)_{L} 
\end{array} \right\} 
L_L  (3, +1)  &
\begin{array}{c}
~ (113\bar3 )_3\supset\tau^-_{R}~ (1_1, -1)   \\
~ (113\bar3 )_2\supset\mu^-_{R} ~ (1_2, -1) \\
~ (113\bar3 )_1\supset e^-_{R} ~ (1_3, -1)  
\end{array}
&
\begin{array}{c}
~and~~~ N^{(1)}_{R} ~ (1_1, +1) \\
~ and~~~N^{(2)}_R ~ (1_2, +1) \\
~ and~~~N^{(3)}_{R} ~ (1_3, +1) \ . \\  
\end{array}
\end{array}
\label{leptons} 
\ee

\bigskip

\noindent
For the left handed quarks  we make the assignment
\begin{equation}
\begin{array}{cc}
(3\bar3 11)_3\supset\left( \begin{array}{c} t \\ b \end{array} \right)_{L}
~ {\cal Q}_L ~~~~~~~~~~~ ({\bf 1_1}, +1)   \\
\left. \begin{array}{c} (3\bar3 11)_2\supset\left( \begin{array}{c} c \\ s \end{array} \right)_{L}
\\
(3\bar3 11)_1\supset\left( \begin{array}{c} u \\ d \end{array} \right)_{L}  \end{array} \right\}
Q_L ~~~~~~~~ ({\bf 2_1}, +1) \ .
\end{array}
\label{qL}
\end{equation}

\noindent Finally, we need assignments for the six right-handed quarks. They were assigned to

\begin{equation}
\begin{array}{c}
t_{R} ~~~~~~~~~~~~~~ ({\bf 1_1}, +1)   \\
b_{R} ~~~~~~~~~~~~~~ ({\bf 1_2}, -1)  \\
\left. \begin{array}{c} c_{R} \\ u_{R} \end{array} \right\}
{\cal C}_R ~~~~~~~~ ({\bf 2_3}, -1)\\
\left. \begin{array}{c} s_{R} \\ d_{R} \end{array} \right\}
{\cal S}_R ~~~~~~~~ ({\bf 2_2}, +1) \ ,
\end{array}
\label{qR}
\end{equation}

\noindent
under $T' \times Z_2$ in Ref.~\cite{Frampton:2008bz} (FKM). However, this assignment 
is inapplicable here as  $t_R$ and $b_R$ are both in the same irrep $(\bar3113)_3$, despite 
having different T' assignments (likewise for the first and second families). Without 
additional states, we are able to assign only three of the six right-handed quarks.

\bigskip

\noindent
We therefore add an anomaly-free sub-quiver representation
\be
3[(\bar{3}, 1, 3, 1)^{'} + (1, 1, \bar{3}, 3)^{'} + (3, 1, 1, \bar{3})^{'}] \ ,
\ee
and reassign {\it all} fermions with $Z_2=-1$, including the corresponding
subset in Eq. (\ref{leptons}) and Eq. (\ref{qR}), to this sub-quiver:
\be
\begin{array}{c}
b_R ~~~~~~~~~~~~~~~~~~~ \subset (\bar{3}, 1, 3, 1)^{'}_3\\
{\cal C}_R ~~~~~~~~~~~~~~~~~~~\subset (\bar{3}, 1, 3, 1)^{'}_{1,2} \\
\tau^-_R ~~~~~~~~~~~~~~~~~ \subset (1, 1, \bar{3}, 3)^{'}_3 \\
\mu^-_R ~~~~~~~~~~~~~~~~~~ \subset (1, 1, \bar{3}, 3)^{'}_2 \\
e^-_R ~~~~~~~~~~~~~~~~~~\subset (1, 1, \bar{3}, 3)^{'}_1 \ .
\end{array}
\label{subquiver}
\ee

\section{Yukawa couplings}

We introduce notation in which the $SU(3)$ groups ($C, R, {\ell}, L$) in 
superscripts are assigned to the fundamental $3$, while those in subscripts are 
assigned to the anti-fundamental $\bar{3}$. The $SU(3)$ groups not denoted in subscript or superscript 
are designated as singlets in this representation. Additionally, the $T^{'}$ assignment will 
be listed in parenthesis with the $Z_2$ charge is given as superscript.

\bigskip

\noindent 
With this stated, the lepton Yukawas are denoted:
\begin{equation}
\Sigma_{i=1}^{i=3} Y_D^{(i)} L^L_{\ell} (3^+) N^{\ell (i)}_R (1_i^+) H_L^R (3^+) 
\label{leptons1}
\end{equation}

\noindent 
and
\begin{eqnarray}
&&\Sigma_{i=1}^{i=3}  Y^{(i)}_{\ell} L_{\ell}^L (3^+) \ell_R^{\ell (i)}(1_i^+)  H_L^R (3^-) \ .
\label{leptons2}
\end{eqnarray}

\bigskip

\noindent

\bigskip

\noindent 
The quark Yukawa couplings are then given as:
\bigskip
\begin{eqnarray}
&& Y_t {\cal Q}_L^C (1_1^+) t^R_C (1_1^+) H_R^L (1_1^+)  + \nonumber \\
&& Y_b {\cal Q}_L^{C} (1_1^+) b_C^{\ell} (1_2^-) H_{\ell}^L (1_3^-) + \nonumber \\
&& Y_{{\cal Q} {\cal S}} {\cal Q}_L^C (1_1^+) {\cal S}_C^R (2_2^+) H_R^L (2_3^+) + \nonumber \\
&& Y_{{\cal C}} Q_L^C (2_1^+){\cal C}_C^{\ell} (2_3^-) H_{\ell}^L (3^-) + \nonumber \\
&&Y_{{\cal S}} Q_L^C (2_1^+)  {\cal S}_C^R (2_2^+) H_R^L (3^+) \ ,
\end{eqnarray}

\bigskip

\noindent
where the $T^{'}$ representations with superscript $Z_2 = +$ are in the original quiver
and all those with superscript $Z_2 = -$ are in the sub-quiver.

\bigskip

\noindent
The Higgs scalar sector is sufficient to break to the standard model and replicate 
the mixing matrices for $T^{'}$ found previously. Note that, for example,
the Cabibbo angle in Ref.~\cite{Frampton:2008bz} follows because after
breaking of $SU(3)_{\ell} \times SU(3)_R$ the $H (3^-) s$ have a common representation,
and can thus act as the appropriate messenger between the charged
leptons and the first two families of quarks. The $T^{'}$ doublet
$(2_3^+)$ of Higgs allows reproduction of the successful CKM
matrix derived in Ref.~\cite{EFM}.

\bigskip

\noindent
The Higgs vacuum expectation values (hereafter VEVs) follow a form highly similar to that in \cite{Frampton:2008bz},
using the same superscript and subscript notation as above. We put the neutral member of the Higgs doublet
at $\alpha_L = 3$ and the corresponding VEV for ($T^{'}=1_1, Z_2=+$) as
\begin{eqnarray}
&&  < H^{L(\alpha_L=3)}_{R(\alpha_R=1)} (1, 3, 1, \bar{3}; 1_1^+) > = \frac{m_t}{Y_t}   \ ,
\label{vev1}
\end{eqnarray}

\bigskip

\noindent
while we put the third family Higgs VEV at $\alpha_R=1$ and the VEV for ($T^{'}=1_3, Z_2=-$) as
\bigskip
\begin{eqnarray}
&&  <H_{\ell (\alpha_{\ell}=1)}^{L(\alpha_L=3)} (1, 3, \bar{3}, 1; 1_3^-) > = \frac{m_b}{Y_b} \ ,
\label{vev2}
\end{eqnarray}

\bigskip

\noindent
with an $\alpha_{\ell}=1$ assignment in the $\ell$-sector. There remain three more VEVs,
which are $T^{'}$ nonsinglets, so we now indicate their direction in $T^{'}$ - space to be:
\bigskip

\begin{eqnarray}
&& <H_{R}^{L} (2_3^+) > ~~~ \propto ~~~ (1,1) 
\label{vev3}
\end{eqnarray}

\begin{eqnarray}
&& <H_{L}^{R} (3^{-}) > ~~~ \propto ~~~ (\frac{m_{\tau}}{Y_{\tau}},\frac{m_{\mu}}{Y_{\mu}},\frac{m_{e}}{Y_{e}}) 
\label{vev4}
\end{eqnarray}

\begin{eqnarray}
&& <H_{L}^{R} (3^{+})> ~~~ \propto ~~~ (1,-2,1)  \ .
\label{vev5}
\end{eqnarray}

\bigskip

\noindent
This collection of five Higgs VEVs can break both the gauge group to
the standard model and achieve the quark and lepton masses as
previously derived in Ref.~\cite{Frampton:2008bz} and elaborated on 
in Refs.~\cite{Frampton:2008ci,EFM1}. In the most general potential 
involving all the scalar fields, there is such a surfeit of parameters that
stationarization of such a potential can, in general, always allow a 
stable global minimum corresponding to the VEVs assumed 
in Eqs.~(\ref{vev1}~-~\ref{vev5}).

\section{Discussion}

We have constructed a consistent quiver unified framework,
based on $(T^{'} \times Z_2)_{global} \times [SU(3)^4]_{local}$
which subsumes the mixing angle predictions for the leptons and quarks
previously made using $T^{'}$ flavor symmetry. Its quiver unification 
predicts additional gauge bosons and bifundamental fermions
at the TeV scale. The production and decay of the lightest Higgs at LHC
can be such as to facilitate discovery
of $H \rightarrow \gamma\gamma$ as was the case in Ref.~\cite{FHKM}.

\bigskip

\noindent
This model illustrates how non-family-sequential flavor symmetry 
$(T^{'} \times Z_2)$, while incompatible with a simple GUT model like 
$SU(5)$, can be wedded successfully to $SU(3)^4$ quiver
unification.

\bigskip

\section*{Acknowledgments}
The work of PHF was supported by DOE grant number DE-FG02-05ER41418;
The work of TWK was supported by DOE grant number DE-FG05-85ER40226; 
they both thank the Aspen Center for Physics for hospitality 
while this work was in progress.

\end{document}